\documentclass[sigconf, nonacm]{acmart}
\begin{document}

\title{koboshi: A Base That Animates Everyday Objects}


\author{Yuta Sugiura}
\affiliation{%
  \institution{Keio University}
  \city{Yokohama}
  \state{Kanagawa}
  \country{Japan}}
\email{sugiura@keio.jp}

\renewcommand{\shortauthors}{Sugiura}

\begin{abstract}
We propose a base-shaped robot named "koboshi" that moves everyday objects. This koboshi has a spherical surface in contact with the floor, and by moving a weight inside using built-in motors, it can rock up and down, and side to side. By placing everyday items on this koboshi, users can impart new movement to otherwise static objects. The koboshi is equipped with sensors to measure its posture, enabling interaction with users. Additionally, it has communication capabilities, allowing multiple units to communicate with each other.
\end{abstract}




\maketitle

\label{sec:scenario2}
\begin{figure}[t]
  \centering
  \includegraphics[width=\linewidth]{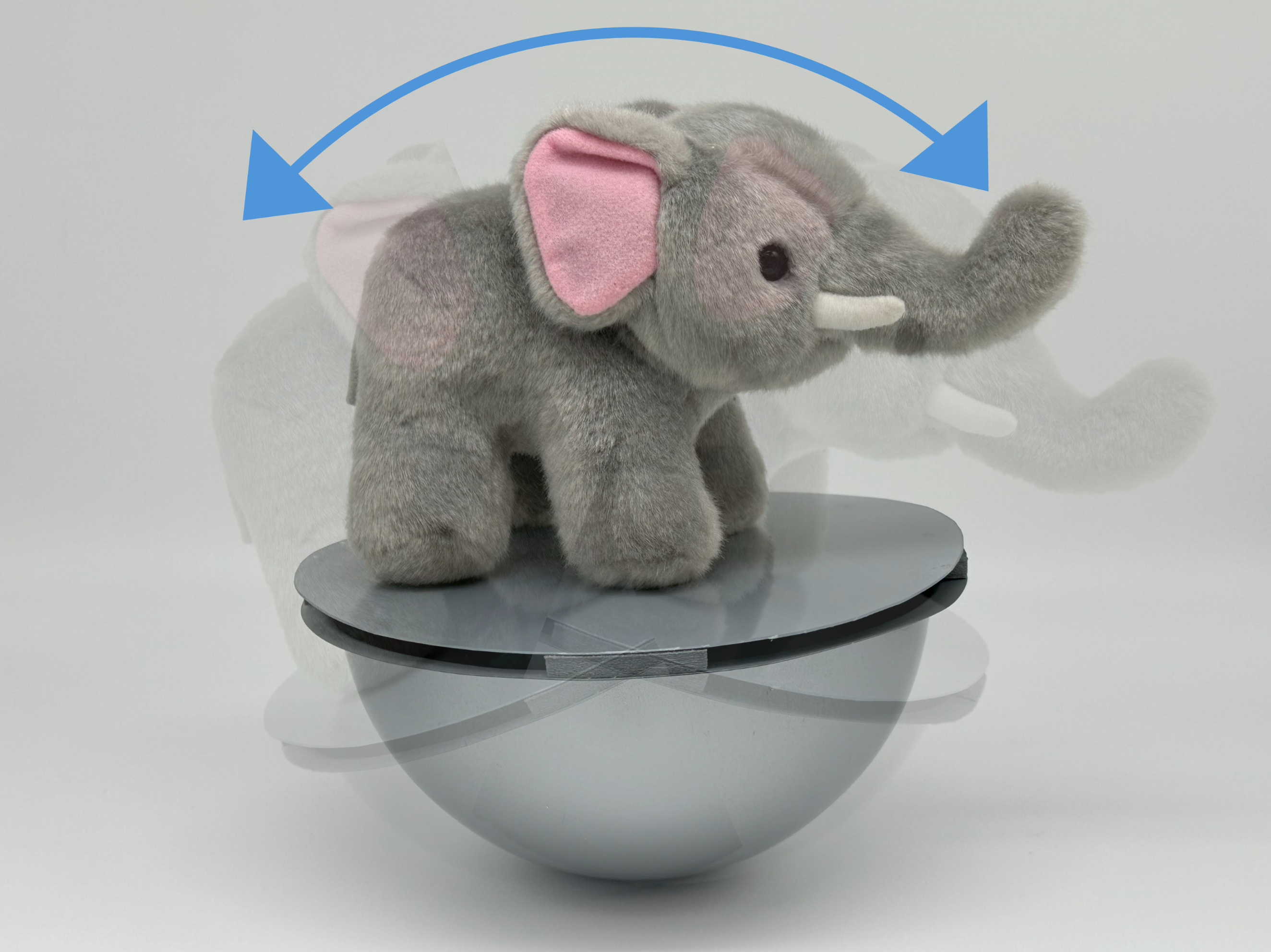}
  \caption{Overview of the koboshi.}
  \label{fig:fig1}
  \Description{System flow.}
\end{figure}

\section{Introduction}
The desire for everyday objects and creations to move has been a common aspiration throughout history. In the movie "Pinocchio," the woodcarver Geppetto wishes for his wooden puppet Pinocchio to come to life, and this wish is magically granted. In the realm of computer graphics, numerous studies focus on imparting motion to still images \cite{As-rigid-as-possible}. However, the desire to animate real-world objects remains a significant challenge.

There have been previous studies on robotizing existing objects, but we propose a novel method to impart motion. One source of inspiration is the traditional Japanese toy called "okiagari-koboshi." This toy has the characteristic of naturally righting itself when knocked over, making it a beloved toy for children. The "okiagari-koboshi" is also known worldwide as the roly-poly toy.

We extend the principle of the roly-poly toy to propose a robotics system that imparts motion to everyday objects. Specifically, by combining this roly-poly mechanism with household items, these objects can start moving as robots (see Figure \ref{fig
}). Additionally, the roly-poly mechanism is equipped with sensors that adjust the center of gravity based on how the user places it, ensuring stable self-righting.

This system allows for interactive contact with users and facilitates communication among multiple units. In this paper, we describe the fundamental structure and operating principles of this system. We also provide a demonstration video of the system\footnote{\url{https://youtu.be/u_7nt4eQ1ns?si=kyKSOg8NGOMgDqd8}}.

\section{Related Work}
\subsection{Robotizing Existing Objects}
In the computer graphics area, numerous studies have focused on imparting motion to animations. As-rigid-as-possible shape manipulation proposes a method to create characters and flexibly generate their animations \cite{As-rigid-as-possible}. However, beyond software, there have been significant advancements in animating real-world objects. Topobo is a modular robotic system that imparts motion to physical structures \cite{topobo}. Additionally, there are studies focused on attaching devices to everyday objects to animate them. For instance, PINOKY is a device that attaches to the tail of a plush toy to give it movement \cite{pinoky}. OmniSkins uses multiple artificial muscles to create devices that can bend the limbs of stuffed animals in various directions \cite{OmniSkins}. Animating Paper achieves bending of paper materials by passing electricity through biomimetal attached to the paper \cite{animatingpaper}. Furthermore, Animated Paper realizes bending by applying heat through a laser to biomimetal actuators attached to paper, eliminating the need for batteries in the object itself \cite{animatedpaper}. PotPet enhances interaction between humans and plants by placing a wheeled robot under a flowerpot \cite{potpet}. Flona is a study that imparts motion to plants by attaching threads and pulling them with appropriate force \cite{flona}. Phones on Wheels develops a wheel module attachable to smartphones to expand human interaction \cite{PhonesonWheels}. SyncPresenter is a highly flexible turntable system that can be taught and replay movements synchronized with sound \cite{SyncPresenter}. Our research provides motion to everyday objects by making the base on which they are placed move. This allows for robotization without significant modifications to the objects.

\subsection{Generating Motion with Weight Shifting}
OMOY is a handheld robotic device that expresses emotions and intentions by moving weights inside it \cite{OMOY}. Building on this concept, a module that adds emotional expression to handheld objects through weight shifting was developed \cite{Weight-ShiftingModule}. Dancer-in-a-box is a system where the box moves as if it is dancing due to the movement of weights inside it \cite{Dancer-in-a-Box}. coconatch is a technology where an object shaped like a coconut moves using internal weights \cite{coconatch}. Our research proposes a method to impart motion to everyday objects using the principle of a roly-poly toy. This approach provides a new functionality of movement to everyday objects, thereby enhancing human-object interaction.

\section{Design}

What koboshi accomplishes is enabling everyday objects to exhibit a swaying motion when placed on its base. (1) The user selects an everyday object to place on koboshi or creates an object they wish to use. (2) They place the object on koboshi. koboshi adjusts its center of gravity according to the placement of the object to remain upright. (3) By moving its internal weights, koboshi performs various motions, providing expressions and interactions to the user.

To achieve the aforementioned scenario, the following design requirements were established for koboshi. (1) It must be able to accommodate everyday objects placed on its base. (2) It must possess intelligence capable of tolerating human imprecision. (3) It must be capable of performing a variety of movements.

\section{Implementation}
\subsection{Hardware}
\begin{figure}[t]
\centering
\includegraphics[width=\linewidth]{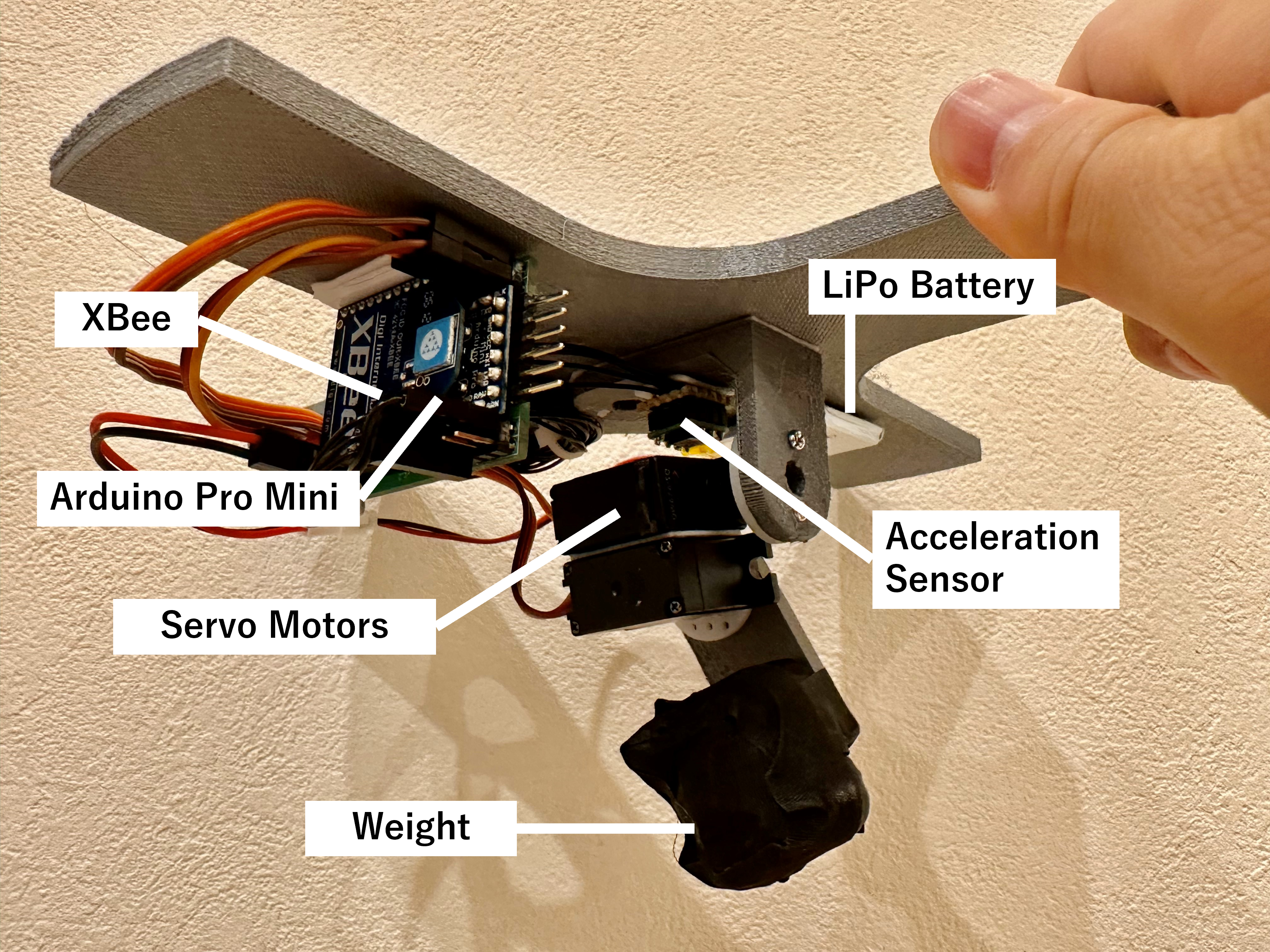}
\caption{The internal appearance of the device.}
\label{device}
\Description{System flow.}
\end{figure}

\begin{figure}[t]
\centering
\includegraphics[width=\linewidth]{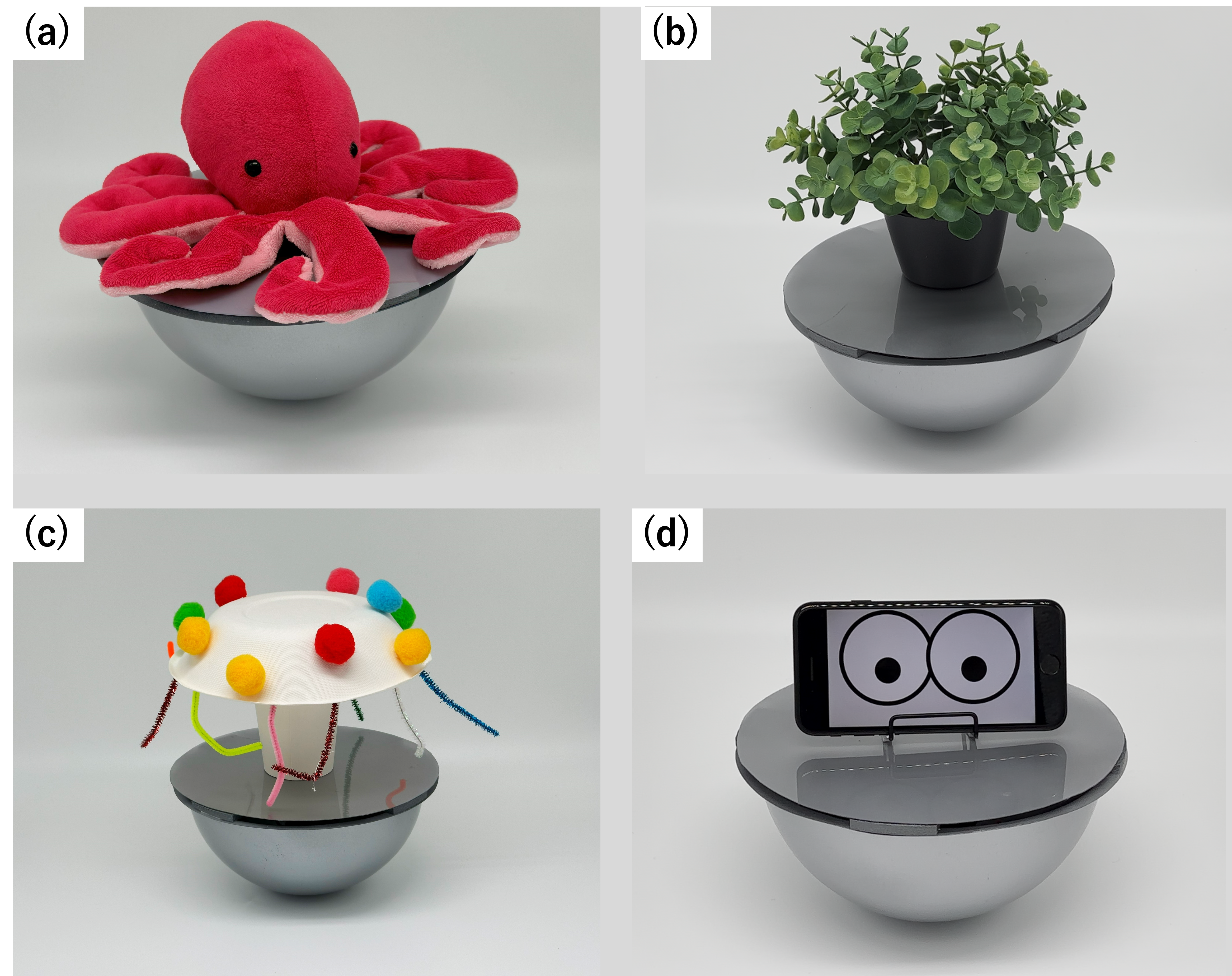}
\caption{Examples of various items placed on koboshi: (a) plush toy, (b) plant, (c) paper craft, (d) smartphone.}
\label{example}
\Description{example.}
\end{figure}

Two servo motors are arranged on orthogonal axes, each connected via arms. At the end of each motor arm, a weight of 20 g is attached, generating motion through their movement (see Figure \ref{device}). Furthermore, the motors are connected to an Arduino Pro Mini microcontroller. This Arduino is powered by a LiPo battery and equipped with XBee, enabling communication between devices and with a PC. An accelerometer is mounted on the top of the device, allowing it to measure its posture. Various everyday objects can be placed on this base \ref{example}.

\subsection{Functions}
\subsubsection{Posture Adjustment}
The servo motor angles are updated to ensure the accelerometer values remain within a specific range. The accelerometer values use the x and y axes to keep the surface of koboshi parallel to the ground.

\subsubsection{Movements}
koboshi is equipped with two servo motors, each operating independently. By controlling the dynamics and speed of these two motors, a variety of movements can be achieved.
\begin{itemize}
\item \textbf{Tilting movements}: The weights tilt in a specific direction and can maintain the tilted position.
\item \textbf{Swaying movements}: The weights move alternately to produce swaying at a specific frequency.
\item \textbf{Vibrating}: The servo motors move at high speed to generate vibrations.
\end{itemize}

\section{Conclusion}
In this study, we proposed a base-shaped robot called "koboshi" that moves everyday objects. We successfully implemented a feature that adjusts the center of gravity to maintain balance according to the position of the objects placed on it. The paper presents examples of using koboshi with various items, such as plush toys, paper crafts, plants, and smartphones. Additionally, we designed interaction features, including remote control and communication between devices. Moving forward, we are interested in exploring how people will engage in creative activities using koboshi. We plan to conduct workshops and other activities to further investigate this aspect.



\bibliographystyle{ACM-Reference-Format}
\bibliography{sample-base}

\end{document}